\documentstyle[12pt,amsfonts]{article}

\def\beq#1{\begin{equation}\label{#1}}
\def\eeq{\end{equation}}
\def\dys{\displaystyle}
\def\mst{\mathstrut}
\renewcommand{\theequation}{\arabic{section}.\arabic{equation}}
\catcode`\@=11 \@addtoreset{equation}{section}\catcode`\@=12

\newcommand{\bear}[1]{\begin{eqnarray}\label{#1}}
\newcommand{\ear}{\end{eqnarray}}

\newcommand{\R}{{\mathbb R}}
\newcommand{\fnm}{\footnotemark}
\newcommand{\fnt}{\footnotetext}
\newcommand{\sign}{ \mbox{\rm sign} }

\newcommand{\sh}{\sinh}
\newcommand{\ch}{\cosh}

\newcommand{\nn}{\nonumber}

\begin{document}

\begin{center}
\large\bf

ON BLACK HOLE  SOLUTIONS \\
IN MODEL WITH ANISOTROPIC FLUID

\vspace{15pt}

%\\[15pt]
\normalsize\bf

H. Dehnen \fnm[1]\fnt[1]{Heinz.Dehnen@uni-konstanz.de},

\it Universit$\ddot{a}$t Konstanz, Fakult$\ddot{a}$t f$\ddot{u}$r Physik,
Fach  M 568, D-78457, Konstanz \\

\vspace{5pt}

\bf V.D. Ivashchuk\fnm[2]\fnt[2]{ivas@rgs.phys.msu.su}
and  V.N. Melnikov \fnm[3]\fnt[3]{melnikov@rgs.phys.msu.su}

\it Center for Gravitation and Fundamental Metrology,
VNIIMS, 3/1 M. Ulyanovoy Str.,
Moscow 119313, Russia  and\\
Institute of Gravitation and Cosmology, PFUR,
Michlukho-Maklaya Str. 6, \\ Moscow 117198, Russia

\end{center}
\vspace{15pt}

\begin{abstract}

A family of spherically symmetric solutions in the model with
1-component anisotropic fluid is considered.  The metric of the solution
depends on a parameter $q > 0$
relating  radial pressure and  the density
and contains  $n -1$ parameters corresponding to
Ricci-flat ``internal space'' metrics.
For  $q = 1$  and certain
equations of state ($p_i = \pm \rho$) the metric  coincides with
the metric of black brane solutions in the model with
antisymmetric form.
A family of black hole solutions corresponding to  natural
numbers $q = 1,2, \ldots$ is singled out.
Certain examples
of solutions (e.g. containing for $q =1$
Reissner-Nordstr\"{o}m,
$M2$ and $M5$ black brane metrics) are considered.
The post-Newtonian parameters $\beta$ and $\gamma$ corresponding to
the 4-dimensional section of the metric are calculated.

\end{abstract}

\pagebreak

%%%%%%%%%%%%%%%%%%%%%%%%%%%%%%%%%%%%%%%%%%%%%%%%%%%%%%%%%%%%%%%%
\section{Introduction}
%%%%%%%%%%%%%%%%%%%%%%%%%%%%%%%%%%%%%%%%%%%%%%%%%%%%%%%%%%%%%%%%

Currently, there is a certain interest to
$p$-brane  solutions with horizon (see, for example,
\cite{IMtop} and references therein) defined on product
manifolds $\R \times M_0 \times \ldots \times M_n$. $P$-brane
solutions (e.g. black brane ones)
usually appear in the models with antisymmetric forms and
scalar fields (see also \cite{St}-\cite{IMp2}).
Cosmological and spherically symmetric solutions with $p$-branes are usually
obtained by the reduction of the field equations to the Lagrange equations
corresponding to Toda-like  systems \cite{IMJ}. An analogous
reduction for the models with multicomponent  "perfect" fluid was
done earlier in \cite{IM5,GIM}.
Earlier extentions of the Schwarzchild, Tangerlini, Reissner-Nordstrom
and Majumdar-Papapetrou solutions to diverse dimensions see in
\cite{Me1,Me2}.

 For cosmological models with
antisymmetric forms without scalar fields any $p$-brane is
equivalent to  an anisotropic perfect fluid with the equations of
state:
\beq{0}
p_i = -   \rho,  \quad {\rm or} \quad  p_i =  \rho,
\eeq
when the manifold $M_i$ belongs or does not belong to the
brane worldvolume, respectively (here $p_i$ is the  pressure
in $M_i$  and $\rho$ is the density, see Section 2).

In  this paper we use this analogy in order to find a new family of exact
spherically-symmetric solutions
in the model with 1-component anisotropic fluid  for more
general  equations of state (see Appendix for more familiar form of
eqs. of state):
\beq{0.1}
p_r = - \rho (2q-1)^{-1}, \qquad p_0 = \rho (2q-1)^{-1},
\eeq
and
\beq{0.2}
p_i= \left(1-\frac{2U_i}{d_i} \right) \rho/(2q-1),
\eeq
$i > 1$,
where $\rho$ is a density, $p_r$ is a radial pressure,
$p_i$ is a pressure in $M_i$, $i = 2, \dots, n$.
Here parameters $U_i$ ($i > 1$) are arbitrary and
the parameter $q > 0$ obey $q \neq 1/2$.
The manifold $M_0$ is $d_0$-dimensiuonal sphere
in our case and $p_0$ is the pressure in the tangent direction.
The case $q = 1$ was considered earlier in \cite{IMS}.

The paper is organized as follows. In Section 2
the model is formulated. In Section 3 a subclass
of spherically symmetric solutions
(generalizing solutions from \cite{IMS}) is presented
and black hole solutions with integer $q$ are singled out.
Section 4  deals with certain examples of solutions
containing for $q =1$
the Reissner-Nordstr\"{o}m metric,
$M 2$ and $M 5$ black brane metrics.
In Section 5  the post-Newtonian
parameters for the 4-dimensional section
of the metric are calculated.

%%%%%%%%%%%%%%%%%%%%%%%%%%%%%%%%%%%%%%%%%%%%%%%%%%%%%%%%%%%%%%%%
\section{The model}
%%%%%%%%%%%%%%%%%%%%%%%%%%%%%%%%%%%%%%%%%%%%%%%%%%%%%%%%%%%%%%%%

Here, we consider a family of spherically symmetric
solutions to Einstein equations with an anisotropic fluid matter source

\beq{1.1} R^M_N - \frac{1}{2}\delta^M_N R = k T^M_N \eeq
defined on the manifold

\beq{1.2}
\begin{array}{rlll}
M = &{\R}_{.} \quad \times &(M_{0}=S^{d_0})\quad \times &(M_1 =
{\R}) \times M_2 \times \ldots \times M_n,\\ &^{radial
    \phantom{p}}_{variable}&\quad^{spherical}_{variables}&\quad^{time}
\end{array}
\eeq with the block-diagonal metrics
\beq{1.2a} ds^2=
e^{2\gamma (u)} du^{2}+\sum^{n}_{i=0} e^{2X^i(u)}
h^{(i)}_{m_in_i} dy^{m_i}dy^{n_i }. \eeq

Here $\R_{.} = (a,b)$ is interval. The manifold $M_i$ with the metric
$h^{(i)}$, $i=1,2,\ldots,n$, is the Ricci-flat space of dimension
$d_{i}$:

\beq{1.3} R_{m_{i}n_{i}}[h^{(i)}]=0, \eeq and $h^{(0)}$ is
standard metric on the unit sphere $S^{d_0}$

\beq{1.4} R_{m_{0}n_{0}}[h^{(0)}]=(d_0-1)h^{(0)}_{m_{0}n_{0}},
\eeq $u$ is radial variable, $\kappa$ is the multidimensional gravitational
constant, $d_1 = 1$ and $h^{(1)} = -dt \otimes dt$.

The  energy-momentum tensor
is adopted in the following form
\beq{1.5a}
(T^{M}_{N})= {\rm diag}(-(2q-1)^{-1} \rho,
(2q-1)^{-1} \rho \delta^{m_{0}}_{k_{0}}, - \rho , p_2
\delta^{m_{2}}_{k_{2}}, \ldots ,  p_n \delta^{m_{n}}_{k_{n}}),
\eeq
where  $q > 0$ and $q \neq 1/2$.
The pressures $p_i$  and
the density $\rho$ obeys the relations
(\ref{0.2}) with arbitrary constants $U_i$, $i >1$.

In what follows we put $\kappa = 1$ for simplicity.

%%%%%%%%%%%%%%%%%%%%%%%%%%%%%%%%%%%%%%%%%%%%%%%%%%%%%%%%%%%%%%%%
\section{Exact solutions }
%%%%%%%%%%%%%%%%%%%%%%%%%%%%%%%%%%%%%%%%%%%%%%%%%%%%%%%%%%%%%%%%

Let us  define
\bear{2.1}
&&1^{o}.\quad U_0 = 0,
\\ \label{2.1a}
&&2^o.\quad U_1 = q,
\\ \label{2.1b}
&&3^o.\quad (U,U)  = U_i G^{ij}U_j,
\ear
where $U = (U_i)$ is $(n+1)$-dimensional vector and
\beq{2.2a}
G^{ij}=\frac{\delta^{ij}}{d_i} + \frac{1}{2-D}
\eeq
are components of the matrix inverse to the matrix of the
minisuperspace metric \cite{IM0,IMZ}
\beq{2.2}
(G_{ij}) = (d_i \delta_{ij} - d_i d_j) \eeq
and
$D=1+\sum\limits_{i=0}^n {d_i}$ is the total dimension.

In our case the scalar product (\ref{2.1b}) reads
\beq{2.1d}
(U,U) = q^2 + \sum_{i=2}^{n} \frac{U_i^2}{d_i}
+ \frac{1}{2-D} \left(q + \sum_{i=2}^{n} U_i \right)^2.
\eeq
It is proved in Appendix that the relation
$1^{o}$ implies  $(U,U) > 0$.

For the equations of state (\ref{0.1}) and (\ref{0.2})
we have obtained the following spherically symmetric
solutions  to the Einstein equations (\ref{1.1})
(see Appendix)
\bear{12}
ds^{2} = J_{0}\left( \frac{dr^{2}}{1-\frac{2\mu}{r^{d}}} +
r^{2} d \Omega^2_{d_0} \right)
- J_1 \left(1-\frac{2\mu}{r^{d}}\right)dt^{2}
\\ \nn + \sum_{i=2}^{n}
J_{i} h^{(i)}_{m_{i}n_{i}} dy^{m_{i}}dy^{n_{i}},
\\ \label{13}
\rho=\dys\frac{\mst (2q-1)(dq)^2
P(P+2\mu)(1 - 2 \mu r^{-d})^{q-1} }{2 (U,U) H^2 J_0 r^{2d_0}},
\ear
by methods similar to obtaining $p$-brane
solution \cite{IMJ}. Here $d=d_0-1$, $d \Omega_{d_0}^2=
h^{(0)}_{m_{0}n_{0}} dy^{m_{0}}dy^{n_{0}}$
is spherical element, the metric factors

\beq{2.3}
J_{i} = H^{-2U^{i}/(U,U)},
\quad  H =
1+ \frac{P}{2 \mu} \left[1 - \left(1 - \frac{2 \mu }{r^{d}}\right)^q
\right];
\eeq $P >0$, $\mu >0$ are constants and

\beq{2.4}
U^{i} = G^{ij}U_{j}  = \frac{U_i}{d_i} + \frac{1}{2-D}
\sum_{j=0}^{n}U_j.
\eeq

Using (\ref{2.4}) and  $U_0 =0$
we get
\beq{2.4a}
U^{0} =  \frac{1}{2-D} \sum_{j=0}^{n}U_j
\eeq
and hence one can rewrite (\ref{12}) as follows
\bear{12a}
ds^{2} = J_{0} \left[ \frac{dr^{2}}{1-\frac{2\mu}{r^{d}}}
+ r^{2} d \Omega^2_{d_0} -
H^{-2q/(U,U)} \left(1-\frac{2\mu}{r^{d}}\right)dt^{2} +
\right.
\\ \nn
\left.
{\qquad} + \sum\limits_{i=2}^{n} H^{-2U_i/(d_i(U,U))}
h^{(i)}_{m_{i}n_{i}} dy^{m_{i}}dy^{n_{i}}
\right].
\ear

{\bf Remark 1.}
We note that the density $\rho$ is positive for
$2q >1$ and negative for $2q < 1$. For $2q  =1$
the solution also exists. In this case $\rho = 0$
and the  energy-momentum tensor should be rewritten
in terms of $p_0$
\beq{1.5b}
(T^{M}_{N})= {\rm diag}(-p_0, p_0 \delta^{m_{0}}_{k_{0}}, 0 , p_2
\delta^{m_{2}}_{k_{2}}, \ldots ,  p_n \delta^{m_{n}}_{k_{n}}),
\eeq
where
\beq{13b}
p_0= \dys\frac{\mst d^2
P(P+2\mu)(1 - 2 \mu r^{-d})^{-1/2} }{8 (U,U) H^2 J_0 r^{2d_0}}.
\eeq

{\bf Black holes for natural $q$.}
For natural
\beq{12n}
q = 1,2, \ldots
\eeq
the metric  has a horizon at $r^d = 2 \mu = r_h^2$.
Indeed, for these values of $q$ the function $H(r) > 0$
is smooth in the interval  $(r_{*}, + \infty)$
for some  $r_{*} < r_h$. (For odd $q = 2m+1$ one get $r_{*} = 0$.
A global structure of the black hole solution corresponding
to these values of $q$ will be a subject of a separate
publication.

For $2 U^0 \neq -1$ and $0< q <1$ we get
a singularity $r^d \to 2 \mu$.
Indeed, due to Einstein equations the scalar curvature of the metric
is proportional to
\beq{12b}
T^M_M = [D -2 -2 \sum_{j=0}^{n}U_j ] p_0 =  (1  + 2 U^0)(D-2) p_0
\eeq
but $p_0$  (proportional to $\rho$)
diverges when $r^d \to 2 \mu$  for $q < 1$ (see (\ref{13})).

{\bf Remark 2.} For non-integer $q >1$ the function $H(r)$ has
a non-analitical behaviour in the vicinity
of $r^d = 2 \mu$.  In this case one may conject
that  the limit $r^d \to 2 \mu$ also corresponds to singularity
but this subject needs a separate investigation.

%%%%%%%%%%%%%%%%%%%%%%%%%%%%%%%%%%%%%%%%%%%%%%%%%%%%%%%%%%%%%%%%
\section{Examples }
%%%%%%%%%%%%%%%%%%%%%%%%%%%%%%%%%%%%%%%%%%%%%%%%%%%%%%%%%%%%%%%%

Up till now $U_i$ were arbitrary in our solution.

Here we consider certain examples of the solution with
\beq{3.0}
U_2 = q d_2 ,\qquad U_i = 0,
\eeq
for $i > 2$ and the equations of state:
$p_2=  - \rho$,
and $p_j=  \rho/(2q-1)$ for $j >2$.
The $q=1$ case describing the
fluid analogue of $p$-brane solution with $p = d_2$
was considered  in \cite{IMS}.

%%%%%%%%%%%%%%%%%%%%%%%%%%%%%%%%%%%%%%%%%%%%%%%%%%%%%%%%%%%%%%%%
\subsection{Solutions for $D=4$}
%%%%%%%%%%%%%%%%%%%%%%%%%%%%%%%%%%%%%%%%%%%%%%%%%%%%%%%%%%%%%%%%

Let us consider the  $4$-dimensional space-time manifold $\R \times
S^2 \times \R$. The metric and the density
>from (\ref{12a}) and (\ref{13}) read
\bear{3.1}
ds^2 = H^{2/q} \left[\frac{ dr^2}{1-\frac{2\mu}{r}} + r^2
d\Omega^2_2  - H^{- 4/q}\left(1-\frac{2\mu}{r}\right) dt^2 \right],\\
\rho = \frac{ (2q-1)P(P+2\mu)}{H^{2+(2/q)} r^4}
\left(1-\frac{2\mu}{r}\right)^{q-1} .
\ear

Here $H = 1+ \frac{P}{2 \mu} \left[1 - \left(1 - \frac{2 \mu }{r}\right)^q
\right]$. For $q = 1$ by changing of variable $r = r' - P$
we obtain the standard Reissner-Nordstr\"{o}m  metric

\beq{3.2}
ds^2_{RN} = -\left(1-\frac{2 GM}{r'}
+\frac{Q^2}{{r'}^2}\right)dt^2 + \left(1-\frac{2 GM}{r'}
+\frac{Q^2}{{r'}^2}\right)^{-1}d{r'}^2 + {r'}^2 d\Omega^2 \eeq
with the charge squared $Q^2 = P(P+2\mu)$ and the gravitational
radius $GM = P+\mu$. Here, obviously, $Q^2 < (GM)^2$.

%%%%%%%%%%%%%%%%%%%%%%%%%%%%%%%%%%%%%%%%%%%%%%%%%%%%%%%%%%%%%%%%
\subsection{Solutions for $D=11$}
%%%%%%%%%%%%%%%%%%%%%%%%%%%%%%%%%%%%%%%%%%%%%%%%%%%%%%%%%%%%%%%%

\quad
Here we consider two examples of solutions
for the case $D = 11$ and $n=3$. These solutions are generalizations
of solutions for $q = 1$ from \cite{IMS}.

{\bf $(M2)_q$-solutions}.
 For $U_2 = q d_2 =2q$, $U_3 = 0$ we get from
(\ref{12a}):
\bear{3.3}
ds^2 = H^{1/(3q)} \left[  \frac{dr^{2}}{1-\frac{2\mu}{r^{d}}}
+ r^{2} d \Omega^2_{d_0}  -H^{-1/q} \left(1-\frac{2\mu}{r^d}\right) dt^2
\right.  \\ \nn \left.
 + H^{-1/q} h^{(2)}_{m_{2}n_{2}} dy^{m_{2}}dy^{n_{2}} +
h^{(3)}_{m_{3}n_{3}} dy^{m_{3}}dy^{n_{3}}
\right].
\ear

For $q=1$ this formula gives the metric of the electric $M2$
black brane solution in 11-dimensional supergravity \cite{CT}.
The  density (\ref{13}) has the following form:

\beq{3.4}
\rho = \frac{ (2q-1) d^2 P(P+2\mu)(1 - 2 \mu r^{-d})^{q-1} }{4 H^{2 +
(1/3q)} r^{2d_0}}.
\eeq

{\bf $(M5)_q$-solution}. Now we put $U_2 =
q d_2 = 5 q$. The metric reads:

\bear{3.5}
ds^2 = H^{2/(3q)} \left[  \frac{dr^{2}}{1-\frac{2\mu}{r^{d}}} +
r^{2} d \Omega^2_{d_0} \right. -
H^{-1/q} \left(1-\frac{2\mu}{r^d}\right) dt^2 +
\\ \nn \left.
H^{-1/q} h^{(2)}_{m_{2}n_{2}} dy^{m_{2}}dy^{n_{2}} +
h^{(3)}_{m_{3}n_{3}} dy^{m_{3}}dy^{n_{3}}
\right],
\ear
and the density is
\beq{3.6}
\rho = \frac{ (2q-1)d^2 P(P+2\mu)(1- 2\mu r^{-d})^{q-1}}{4 H^{2 +
(2/3q)} r^{2d_0}}.
\eeq

For $q=1$ we get the metric of the magnetic $M5$
black brane solution in 11-dimensional supergravity \cite{CT}.

%%%%%%%%%%%%%%%%%%%%%%%%%%%%%%%%%%%%%%%%%%%%%%%%%%%%%%%%%%%%%%%%
\section{Physical parameters}
%%%%%%%%%%%%%%%%%%%%%%%%%%%%%%%%%%%%%%%%%%%%%%%%%%%%%%%%%%%%%%%%

\subsection{Gravitational mass and PPN parameters}

Here we put $d_0  =2 \ (d =1)$. Let us
consider the 4-dimensional space-time section of the metric
(\ref{12a}). Introducing a new radial variable by the relation:

\beq{3.7} r = R \left(1 + \frac{\mu}{2 R}\right)^2, \eeq
we rewrite the $4$-section in the following form:

\bear{3.8}
ds^2_{(4)} =
H^{- 2 U^0/(U,U)} \left[ - H^{- 2q /(U,U) }
\left(\frac{1-\frac{\mu}{2 R}}{1+\frac{\mu}{2 R}}\right)^2 dt^2
+ \left( 1+\frac{\mu}{2 R} \right)^4 \delta_{ij}
dx^i dx^j \right]
\ear
$i,j = 1,2,3$. Here $R^2 = \delta_{ij} x^i x^j$.

The parametrized post-Newtonian (Eddington) parameters are
defined by the well-known relations

\bear{3.9}
g^{(4)}_{00} = - (1-2V+2\beta V^2) + O (V^3), \\
\label{3.10} g^{(4)}_{ij} = \delta_{ij} (1 + 2 \gamma V) + O
(V^2),
\ear
$i,j = 1,2,3$. Here
\beq{3.11}
V=\frac{GM}{R}
\eeq
is
the Newtonian potential, $M$ is the gravitational mass and $G$ is
the gravitational constant.

>From (\ref{3.8})-(\ref{3.10}) we obtain:
\beq{3.12}
GM = \mu +  \frac{P q(q +U^0)}{(U,U)} \eeq
and
\bear{3.13}
\beta - 1= \frac{|A|}{(GM)^2} (q + U^0), \\
\label{3.13a}
\gamma - 1= - \frac{ P q}{(U,U) GM} (q + 2U^0),
\ear
where \\

$|A| = \frac12 q^2 P (P + 2 \mu)/(U,U)$  (see Appendix),
or, equivalently, \\

$P = - \mu + \sqrt{\mu^2 + 2 |A| (U,U) q^{-2}} > 0$.\\

For fixed $U_i$
the parameter $\beta -1$ is proportional to the  ratio of two physical
parameters:
the anisotropic fluid density parameter $|A|$
and the gravitational radius squared
$(GM)^2$. For compact internal spaces the parameter $|A|$ is proportional
to the effective mass of the fluid outside the external horizon
(for natural $q$), i.e.  to
the integral of $\rho$ over the region $r^d > 2 \mu$.

\subsection{Hawking temperature}

The Hawking temperature of the black hole may be calculated
using the well-known relation \cite{York}
\beq{Y}
      T_H  = \frac{1}{4 \pi \sqrt{-g_{tt} g_{rr}}}
      \frac{d(-g_{tt})}{dr} \Biggl|_{_{\ horizon}}.  \eeq
We get
\beq{H}
T_H =
\frac{d}{4 \pi (2 \mu)^{1/d}}
\left(1 + \frac{P}{2\mu}\right)^{- q/(U,U)}.
\eeq
Here $q = 1,2, \ldots.$

For the $4$-dimensional solution (\ref{3.1}) we get
$T_H = \frac{1}{8 \pi \mu }
\left(1 + \frac{P}{2\mu}\right)^{- 2/q}$.
For $D=11$ metrics (\ref{3.3}) and (\ref{3.5})
the Hawking temperature reads
\\

$T_H =
\frac{d}{4 \pi (2 \mu)^{1/d}}
\left(1 + \frac{P}{2\mu}\right)^{- 1/(2q)}$.
\\

\section{Conclusions}

In this paper, using our methods developed earlier for obtaining perfect
fluid and p-brane solutions, we have considered a family of spherically
symmetric solutions  in the model with 1-component anisotropic fluid when
the equations of state (\ref{0.1}) and  (\ref{0.2}) are imposed.  The
metric of any solution  contains  $(n -1)$ Ricci-flat "internal" space
metrics and depends upon arbitrary parameters $U_i$, $i > 1$.

For $q = 1$ and  certain equations of state (with $ p_i = \pm
\rho$) the metric of the solution  coincides with that of black brane (or
black hole) solution in the model with antisymmetric forms without
dilatons \cite{IMS}.
For natural numbers $q = 1,2, \ldots$  we obtained
a family of black hole solutions.

Here we also considered
certain examples of  solutions with horizon, e.g.
(fluid) generalizations of charged black hole and
$M2$, $M5$  black brane solutions.

We  have also calculated for  possible estimations of observable efffects
of extra dimensions the post-Newtonian parameters $\beta$ and $\gamma$
corresponding to the 4-dimensional section of the metric and the Hawking
temperature as well.  The parameter $\beta -1$ is written in terms of
ratios of the physical parameters: the perfect fluid parameter $|A|$ and
the gravitational radius squared $(GM)^2$.

{\bf Acknowlegments}

This work was supported in part by the Russian Ministry of
Science and Technology, Russian Foundation for Basic Research
(RFFI-01-02-17312-a)and DFG Project (436 RUS 113/678/0-1(R)).

V.D.I. thanks colleagues from the Physical Department of
the University of Konstanz for their
hospitality during his visit in May-July and
V.N.M. - during his visit in Sepember-October, 2002.
We also thank K.A. Bronnikov for fruitful discussions of the paper.

\newpage

\renewcommand{\theequation}{\Alph{subsection}.\arabic{equation}}
\renewcommand{\thesection}{}
\renewcommand{\thesubsection}{\Alph{subsection}}
\setcounter{section}{0}

%%%%%%%%%%%%%%%%%%%%%%%%%%%%%%%%%%%%%%%%%%%%%%%%%%%%%%%%%%%%%%%%
\section{Appendix}
%%%%%%%%%%%%%%%%%%%%%%%%%%%%%%%%%%%%%%%%%%%%%%%%%%%%%%%%%%%%%%%%

%\renewcommand{\theequation}{\Alph{section}.\arabic{equation}}
%\renewcommand{\thesubsection}{\Alph{section}}
%\setcounter{section}{1}

%%%%%%%%%%%%%%%%%%%%%%%%%%%%%%%%%%%%%%%%%%%%%%%%%%%%%%%%%%%%%%%%
\subsection{Lagrange representation}
%%%%%%%%%%%%%%%%%%%%%%%%%%%%%%%%%%%%%%%%%%%%%%%%%%%%%%%%%%%%%%%%

It is more convenient for finding
of exact solutions, to write the stress-energy tensor
in cosmological-type form
\beq{1.5} (T^{M}_{N})= {\rm diag}(-{\hat{\rho}},{\hat p}_{0}
\delta^{m_{0}}_{k_{0}}, {\hat p}_{1}
\delta^{m_{1}}_{k_{1}},\ldots , {\hat p}_n
\delta^{m_{n}}_{k_{n}}),
\eeq
where $\hat{\rho}$ and $\hat p_{i}$
are "effective"  density and pressures, respectively, depending
upon the radial variable $u$ and
the physical density $\rho$
and pressures $p_i$ are related to the effective
("hat") ones by formulas
\beq{1.7a} \rho = - {\hat p}_1, \quad p_r = - \hat{\rho}, \quad
p_i = \hat{p}_i, \quad (i \neq 1).
\eeq

The equations of state may be written in the following
form
\beq{1.7}
{\hat p}_i=\left(1-\frac{2U_i}{d_i}\right){\hat{\rho}},
\eeq
where $U_i$ are constants, $i= 0,1,2,\ldots,n$.
It follows from (\ref{1.7a}), (\ref{1.7}) and $U_1 =q$
that
\beq{1.7b}
\rho = (2q - 1) \hat{\rho}.
\eeq

The conservation law equations
$\nabla_{M} T^M_N = 0$ (following from Einstein equations)
may be written,
due to relations (\ref{1.2a}) and (\ref{1.5}) in the following form:

\beq{5.7} \dot{\hat{\rho}} +\sum_{i=0}^n
d_i\dot{X^i}({\hat{\rho}} +{\hat p}_i )=0. \eeq
Using the equation of state (\ref{1.7}) we get

\beq{5.7a}
{\hat{\rho}}= - A e^{2U_iX^{i}-2\gamma_{0}},
\eeq
where $\gamma_0(X)= \sum\limits_{i=0}^{n} d_{i}X^{i}$ and $A$
is constant.

The Einstein equations (\ref{1.1})  with the relations
(\ref{1.7}) and (\ref{5.7a}) imposed are equivalent to the
Lagrange equations for the Lagrangian

\beq{} L = \frac
{1}{2}e^{-\gamma+\gamma_0(X)}G_{ij}\dot{X}^{i}\dot{X}^{j}
-e^{\gamma-\gamma_0(X)}V, \eeq
where

\beq{5.32n}
V= \frac{1}{2} d_0 (d_0 -1) e^{2U^{(0)}_i X^i} +
A e^{2 U_iX^i}
\eeq
is the potential and the components
of the minisupermetric $G_{ij}$ are defined in (\ref{2.2}).
\beq{5.8}
U^{(0)}_i X^i = -X^0 + \gamma_0(X), \qquad U^{(0)}_i  =
- \delta^0_i + d_i, \qquad A_{0} = \frac{1}{2} d_0 (d_0 -1),
\eeq
$i = 0, \ldots, n$ (for cosmological case see \cite{IM5,GIM}).

For $\gamma=\gamma_0(X)$, i.e. when the harmonic time gauge
is considered, we get the set of Lagrange equations
for the Lagrangian
\beq{5.31n}
L=\frac12G_{ij} \dot X^i \dot X^j-V,
\eeq
with the zero-energy constraint imposed
\beq{5.33n}
E=\frac12G_{ij} \dot X^i \dot X^j + V =0.
\eeq

It follows from the restriction
$U_0 = 0$ that
\beq{5.43a}
(U^{(0)},U)  \equiv U^{(0)}_i G^{ij}U_j = 0.
\eeq

Indeed, the contravariant components $U^{(0)i}=
G^{ij} U^{(0)} j$ are the following ones
\beq{5.43b}
U^{(0)i}=-\frac{\delta_0^i}{d_0}.
\eeq

Then we get $(U^{(0)},U)  = U^{(0)i} U_i = - U_0/d_0 =0$.
In what follows we also use the formula
\beq{5.43c}
(U^{(0)},U^{(0)})   = \frac{1}{d_0} - 1 < 0,
\eeq
for $d_0 >1$.

Now we prove that $(U,U) > 0$. Indeed, minisupermetric
has the signature $(-,+,\ldots,+)$ \cite{IM0,IMZ}, vector $U^{(0)}$
is time-like and orthogonal to vector $U \neq 0$. Hence
the vector $U$ is space-like.

%Thus, the curvature part of potential  may be simulated by the
%fluid with the parameter $U^{(0)}$.

%\addtocounter{section}{1} \setcounter{equation}{0}
\subsection{General spherically symmetric solutions}

When the orthogonality relations  (\ref{5.43a}) and
$3^o$ of (\ref{2.1}) are satisfied the Euler-Lagrange equations for the
Lagrangian (\ref{5.31n}) with the potential (\ref{5.32n}) have the
following solutions (see relations from \cite{GIM} adopted for our
case):

\beq{5.34n}
X^i(u)= -
\sum_{\alpha=0}^1
\frac{U^{(\alpha)i}}{(U^{(\alpha)},U^{(\alpha)})}\ln
|f_{\alpha}(u-u_{\alpha})| + c^i u + \bar{c}^i, \eeq
where
$U^{(1)} = U$, $u_{\alpha}$
are integration constants; and vectors $c=(c^i)$ and $\bar c=(\bar
c^i)$ are orthogonal to the $U^{(\alpha)}=(U^{(\alpha)i})$, i.e. they
satisfy the linear constraint relations

\bear{5.47n}
U^{(0)}(c)= U^{(0)}_i c^i = -c^0+\sum_{j=0}^n d_j c^j=0, \\
\label{5.48n} U^{(0)}(\bar c)= U^{(0)}_i \bar c^i = -\bar
c^0+\sum_{j=0}^n d_j \bar c^j=0, \\
\label{5.49n} U(c)= U_i c^i=0,\\
\label{5.50n} U(\bar c)=  U_i \bar c^i=0. \ear

Here
\beq{A.7}
\begin{array}{rlll}
f_{\alpha}(\tau)=
 & R_{\alpha} \dys\frac{\mst
 \sh(\dys\sqrt{\mst C_{\alpha}}\tau)}{\dys\sqrt{\mst C_{\alpha}}},
 & C_{\alpha} > 0,&
\eta_{\alpha} = +1 ,
\\  & R_{\alpha} \dys\frac{\mst\ch(\dys\sqrt{\mst
C_{\alpha}}\tau)}{\dys\sqrt{\mst C_{\alpha}}},& C_{\alpha}>0,
& \eta_{\alpha} = -1 ,
\\  & R_{\alpha} \dys\frac{\mst\sin(\dys\sqrt{\mst
|C_{\alpha}|}\tau)}{\dys\sqrt{\mst |C_{\alpha}|}},& C_{\alpha}<0,
& \eta_{\alpha} = + 1 , \\\\
 & R_{\alpha} \tau,& C_{\alpha}=0,&
 \eta_{\alpha} = + 1 ,
\end{array}
\eeq
$\alpha = 0,1$; where $R_0 = d_0-1$, $\eta_0 = 1$,
$R_1 =\sqrt{2|A|(U,U)}$, $\eta_1 = - \sign A$.

The zero-energy constraint, corresponding to the solution (\ref{5.34n})
reads

\beq{A.17}
E = \frac12 \sum_{\alpha = 0}^1
\frac{C_{\alpha}}{(U^{(\alpha)} , U^{(\alpha)})} + \frac12
G_{ij}c^ic^j=0 . \eeq

{\bf Special solutions.}
The horizon condition (i.e. infinite
time of propagation of light for $u \to +\infty$) lead us to the
following integration constants
\bear{5.67}
\bar{c}^i  & = &  0,\\
c^i & = & \bar{\mu}  \sum_{\alpha = 0}^1
\frac{U^{(\alpha)}_1 U^{{(\alpha)}i}}{(U^{(\alpha)},U^{(\alpha)})} -
\bar{\mu} \delta^i_1, \\
       \label{5.68}
C_{\alpha}  & = & (U^{(\alpha)}_1)^2  \bar{\mu}^2,
\ear
where $\bar{\mu} > 0$, $\alpha=0,1$.

We also introduce a new radial variable $r = r(u)$ by relations

\beq{5.69}
\exp( - 2\bar{\mu} u) = 1 - \frac{2\mu}{r^d},  \quad
\mu = \bar{\mu}/d >0, \quad  d = d_0 -1,
\eeq
and put $u_1 < 0$, $A  < 0 $,  $u_0 = 0$.

The relations of the Appendix imply the formulae
(\ref{12}) and (\ref{13})
for the solution from Section 3 with
\beq{5.70}
H =  \exp(- \bar{\mu} U_1 u) f_1(u - u_1), \qquad
A = - \frac{(dq)^2}{2(U,U)} P ( P +2 \mu),
\eeq
$P > 0$.

\end{document}